# Electroosmotically enabled Electrorheological Effects in a Planar Nematic Crystal Flow

Jayabrata Dhar, Antarip Poddar and Suman Chakraborty
Department of Mechanical Engineering, Indian Institute of Technology, Kharagpur-721302, India

**Abstract**

Study of electrokinetics of nematic liquid crystals (LCs) with dissolved impurities hold utmost importance in understanding director distribution characteristics and modified flow rheology. However, no concrete theory for the non-uniform potential and ionic species distribution, due to an induced electrical double layer (EDL) at the LC-substrate interface, derived from fundamental principles have been put forward in this regard. In this work, we have developed coupled governing equations from fundamental free energy considerations for the potential distribution and the director configuration of the nematic LC within the induced electrical double layer which is generated due to certain physico-chemical interactions at the LC-substrate interface. With these considerations, an electroosmotically-enabled nematodynamics for a particular LC, namely, MBBA, with strong planar anchoring at the boundaries is studied. We obtained multiple solution for director configuration, which is an integral characteristics of nematic flow solutions, and investigated for the most stable solution employing an entropic analysis. We finally proceed to depicts the electroosmotic flow features of the nematic LC wherein we focused on the spontaneous development of the electrorheological effect and the resulting elastic description of the nematic LC for the most stable solution obtained.

___________________________
email for correspondence: suman@mech.iitkgp.ernet.in

# Introduction

Studies on electro-nematodynamics of liquid crystals is a significant topic in present research due to its diversified applications in display devices [1,2], electro-optical devices [1,3], electrorheological performance [4–8] and electrokinetic flow control [9,10]. Liquid crystals are a particular class of fluids showing long-range orientational or positional order and elastic property due to its inherent crystalline nature of the constituting molecules which may exhibit different shapes [11,12]. Nematics are a sub-class of such liquid crystalline materials generally displaying a rod-shaped molecular structure with an orientational long-order arrangement [11–13]. A directional property is generally connected with the long-axis of these molecules and their average direction is denoted by a unit vector **n** known as the director [11,12]. The orientation and arrangement of the shape-specific molecules may be influenced by various factors including externally applied electric or magnetic field [11,14], externally imposed flow [11,13,15,16] and presence of ionic entities [17,18] in the liquid sample, which in turn, greatly affects the fluid rheology [5,19,20].

A key aspect in any generic nematic cell is the presence of ionic impurities [21–25] which, under the influence of an external electric field, illustrates a distribution pattern within the liquid in order to screen off the applied field, and thereby, alters the resulting flow dynamics [18,26]. The presence of such impurities is in fact a hindrance to the desired effect that the nematic LCs are used for in applications relating to LC displays [9,13,21]. However, recent experimental studies have demonstrated the use of these impurities in the purview of electrokinetic-enabled nematic flows [9,10]. The genesis of such flows is the induction of an electrical double layer (EDL) adjoining the nematic-substrate interface [27–29]. An EDL, which may form due to certain ionic adsorption or physico-chemical mechanisms, comprises of a layer of immobilized charge, known as the Stern layer, directly adjacent to the wall substrate and a layer of mobile ions, known as the Diffuse layer [30,31]. The ions having opposite charge as that of the wall is known as counterions while those having similar charge are known as the coions. Shah and Abbott [27] experimentally verified that the structure of these induced EDLs within nematic liquid crystal has substantial influence on the director orientation near the surfaces. The studies regarding electrokinetics of nematic fluids allow electrical flow actuation that has encouraging consequences on resulting nematodynamics, although no detailed corresponding theoretical foundation governing such flows has been proposed.

In this present study, we address the lack of such a fundamental theoretical basis of a general eletrokinetic flow to bridge the two gaps providing a coupled governing equation for electric double layer potential distribution and electrically actuated sustained flow of nematic LCs. Here we consider an electroosmotic flow of a nematic liquid and explore the electrokinetic and electrorheological effects induced due to such flow actuation. Electroosmosis refers to the advection of diffused excess ionic entity near a charged substrate, which, in turn, drags the fluid along with it resulting in a bulk motion, due to the action of an external axial electric field [30,31] while electrorheology refers to the enhanced viscosity the fluid exhibits in presence of an electric field [32–34]. An interesting display of electrorheological effects have previously been observed for a confined nematic fluid with shear or pressure driven flow actuation and in presence of a transverse electric field [4,6,8,20,35]; however no parallel investigation has been done to explore the rheological variations in presence of an induced EDL with electrical flow actuation. We begin with the fundamental free energy consideration for a nematic sample sandwiched between two parallel plates wherein we assume that an EDL gets established at the nematic-substrate interface generating a transverse non-uniform electric field. Here we have specifically assumed that the conductive anisotropy of the medium is negligible and the consequent charge separation dynamics, known as the Carr-Helfrich effect [13], does not take place. This assumption is consistent with a series of works [26,36–38] which explicitly considered the ionic presence within nematic phase without the consideration of conductive anisotropy driven charge separation. These considerations must be sharply distinguished from non-linear electroosmosis [9,39], where the genesis of the flow is due to the electric field-induced charge separation, and therefore, allows us to focus the present study within the realm of linear electroosmosis. Owing to the induced and the applied electric field, dielectric electrostatic energy [11,13] and flexoelectric energy [13,40–42] are necessarily to be considered besides the elastic Frank-Oseen energy [11,13] for the nematic phase. The minimization of the free energy yields the modified Poisson-Boltzmann equation for potential distribution within the nematic domain [43] while maintaining the classical Boltzmann distribution for the ions in the transverse direction. Furthermore, the excluded volume effects of the ionic shells have also been accounted for in the present formalism. We employ the Leslie-Ericksen formulations for nematic LCs

[11,13] to construct the governing equations for the director distribution and electroosmotic flow within the narrow confinement. Two pivotal aspects of any electrokinetic modulated flow of a nematic, namely, the electrorheological behavior and the normal stress difference have been elaborated in this study.

## Mathematical Formulation

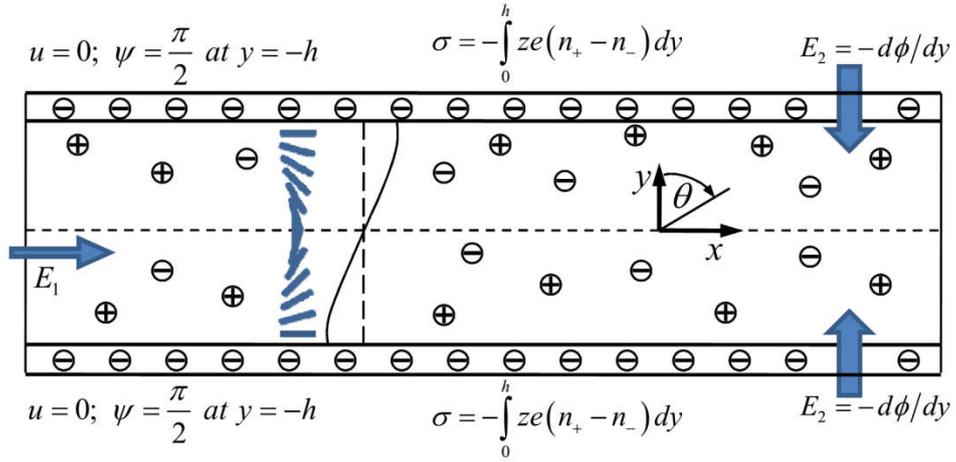

**Fig 1:** Schematics of the electroosmotic flow condition and director deformation of a Nematic liquid crystal phase under the application of an external electric field. The boundary conditions for the flow, director alignment and induced surface charge has been depicted along with the directions of the induced transverse field due to the EDL formation and applied field that actuates the fluid motion.

We consider the electric field driven flow actuation of an elastic nematic liquid crystal, considering splay, twist and bend elastic energies, confined within a slit type flow passage of thickness $2h$ as shown in figure 1. A dielectric anisotropy is associated with the nematic phase having parallel and perpendicular dielectric constant as $\varepsilon_\parallel$ and $\varepsilon_\perp$, respectively, while charged impurities with a number density $n_0$ is assumed to be suspended within the nematic sample. The solid bounding surfaces are identical with a characteristic charge-selective adsorption at the nematic-substrate interface inducing a surface charge density $\sigma_w$. Owing to the induced charge at the interface, ionic charge distribution gets established with the formation of the electrical double layer, which spontaneously induces a transverse electric field. Unlike the case of electroosmosis in Newtonian fluids, this transverse field has an intricate effect on the flow rheology besides affecting the body force generation for the flow actuation. For nematic liquids, the spatially varying unit director vector $\mathbf{n}$, representing the average direction of the nematic molecules, can be simplified in our case as $\mathbf{n} = \sin\theta(y)\hat{\mathbf{i}} + \cos\theta(y)\hat{\mathbf{j}}$ while a planar arrangement with strong anchoring of the director (the director angle is pre-defined at the boundaries [11]) is assumed at both the walls. Upon the application of an external longitudinal electric field, a bulk fluid motion sets in due to electroosmosis wherein the velocity is assumed to be only a function of the transverse direction $\mathbf{V} = u(y)\hat{\mathbf{i}}$. The electroosmotic flow of the nematic not only affects the director deformation and flow rheology but, in sharp contrast to Newtonian case, also have an intrinsically subtle effect on the potential distribution within the EDL which will be subsequently discussed.

**Equilibrium Nemato-static Configuration**

In order to determine the equilibrium director configuration and charge distribution due to the induced EDL within the nematic phase, we begin by considering the total free energy $F$ which comprises of the nemato-elastic energy, the dielectric anisotropic energy, the gradient flexoelectric energy, the internal energy and the entropic contributions that accounts for the excluded volume effects of the dissolved ions.

*Free energy due to elastic distortion*

For the present study, we consider MBBA nematic liquid having rod-shaped molecular structure and assuming planar deformation wherein the director deformations are restricted to the plane of electric flux. Generally, the elastic distortion energy density, consisting of the splay (splay elastic modulus $K_{11}$), twist (twist elastic modulus $K_{22}$) and bend (bend elastic modulus $K_{33}$) energy contributions, is given by the Frank-Oseen formulation [11,13] as

$$F_{elas} = \frac{1}{2}\int \left( K_{11}(\nabla \cdot \mathbf{n})^2 + K_{22}(\mathbf{n} \cdot \nabla \times \mathbf{n})^2 + K_{33}(\mathbf{n} \times \nabla \times \mathbf{n})^2 \right) dy \qquad (1)$$

*Free energy in presence of electric fields*

In the presence of an electric field, the energy associated with the nematic phase having dissolved ionic impurities has contributions from the liquid dielectric anisotropy ($F_{de}$), flexoelectric polarization energy $F_{fe}$ and internal energy $F_{int}$ due to the presence of free ions as

$$F_{el} = F_{de} + F_{flex} + F_{int} \tag{2}$$

The dielectric energy is classically given by $F_{de} = -\frac{1}{2}\int \mathbf{D}\cdot\mathbf{E}\, dV$ where $\mathbf{D}$ is the electric displacement vector and $\mathbf{E} = E_1\hat{\mathbf{i}} + E_2(y)\hat{\mathbf{j}}$ is the electric field vector, where $E_1$ denotes the applied axial field and $E_2 = -\nabla\psi(y)$ denotes the spontaneously induced inhomogeneous transverse field and $\psi(y)$ denotes the potential distribution within the EDL. The electric displacement vector for an anisotropic ordered fluid with orientational order represented by the director $\mathbf{n}$ has the form $\varepsilon_0\left[\varepsilon_\perp \mathbf{E} + \varepsilon_a(\mathbf{E}\cdot\mathbf{n})\mathbf{n}\right]$ where $\varepsilon_a = \varepsilon_\parallel - \varepsilon_\perp$ is known as the dielectric anisotropy and $\varepsilon_0$ is the absolute permittivity of free space. Employing the component form for the director $\mathbf{n}$ and the electric field vector $\mathbf{E}$, the form for the dielectric anisotropic energy density reads

$$F_{de} = \int -\frac{\varepsilon_0 \varepsilon_a}{2}\left[E_1\sin(\theta) + E_2(y)\cos(\theta)\right]^2 - \frac{\varepsilon_0 \varepsilon_\perp}{8\pi}\left[E_1^2 + E_2(y)^2\right] dy \tag{3}$$

In addition to the dielectric energy, a flexoelectric energy is associated with the nematic molecules attributable to their combined induced polarization and structural elastic bending characteristics, which is determined as $F_{flex} = -\int \mathbf{P}_{fl}\cdot\mathbf{E}\, dV$. The induced polarization for such ordered nematic is given by the form [44] $\mathbf{P}_{fl} = e_1(\hat{\mathbf{n}}\nabla\cdot\hat{\mathbf{n}}) + e_3(\hat{\mathbf{n}}\times\nabla\times\hat{\mathbf{n}})$, resulting in the energy density form

$$F_{flex} = \int\left(\left[\left(e_1\sin^2(\theta) + e_3\cos^2(\theta)\right)E_1 + (e_1 - e_3)\sin(\theta)\cos(\theta)E_2(y)\right]\frac{d\theta}{dy}\right)dy \tag{4}$$

where $e_1$ and $e_3$ are the flexoelectric coefficients. Presence of ionic entities within the nematic sample further give rise to the internal energy density which comprises of the self-energy and electrostatic energy given by the form [43]

$$F_{\text{int}} = \int \left( -\frac{\varepsilon_0 \varepsilon_\perp}{2} \left( \nabla \phi(x,y) \right)^2 + ze\phi(x,y)\left( n_+ - n_- \right) \right) dy \tag{5}$$

where we have considered a symmetric $z:z$ electrolyte and $\phi(x,y) = \psi(y) + (\phi_0 - xE_1)$ signifies the total potential due to the combined applied and induced electric field.

*Entropic contributions to free energy*

Relaxing the ideal gas considerations in the entropic contribution to the free energy, we go beyond the point charge approximation for the ionic species accounting for the ion-ion steric interactions due to their excluded volume effects. The free energy density associated with the entropic contributions considering the finite ionic shell size is given by the form [45,46]

$$\begin{aligned}F_{\text{entropic}} = -TS = k_B T \int dy \left[ n_+ \ln\left(a_+^3 n_+\right) + n_- \ln\left(a_-^3 n_-\right) - n_+ - n_- \right] \\ + \frac{k_B T}{a^3} \int dy \left[ \left(1 - a_+^3 n_+ - a_-^3 n_-\right) \ln\left(1 - a_+^3 n_+ - a_-^3 n_-\right) \right]\end{aligned} \tag{6}$$

where $n_+ (n_-)$ refers to the number density of positive (negative) ions and $a_+ (a_-)$ denotes their corresponding ionic shell size. For the sake of simplicity, we assume $a = a_+ = a_-$. It must be noted that employing such a form restricts excessive ionic crowding near the wall which comes into prominence for situations involving concentrated solutions, high surface charge and large sized molecules. With the consideration of the individual contributing energy terms, we have the total free energy for the equilibrium nematic orientation and potential distribution as following:

$$F = F_{\text{elast}} + (F_{de} + F_{flex} + F_{\text{int}}) + F_{\text{entropic}} = \int f_{tot} \, dy \tag{7}$$

where $f_{tot}$ denotes the integrand of the above integral form of the free energy.

*Equilibrium Potential Distribution*

Potential distribution within the EDL for a nematic liquid is largely influenced by its director distribution attributable to its order and dielectric anisotropic characteristic of such fluids [47]. The intricate behavioral condition governing the distribution of potential and the ionic species within the liquid phase may be obtained by minimizing the total free energy with respect to the electrostatic potential in the equilibrium condition. Towards this, we proceed to obtain the modified Poisson-type equation for nematic phase using $\frac{\delta f_{tot}}{\delta \psi} = 0$ to obtain the form

$$\varepsilon_0(\varepsilon_a \cos^2(\theta) + \varepsilon_\perp)\frac{d^2\psi}{dy^2} - \varepsilon_0\varepsilon_a\left(E_1\cos(2\theta) + \sin(2\theta)\frac{d\psi}{dy}\right)\frac{d\theta}{dy} + \frac{1}{2}(e_1 - e_3)\sin(2\theta)\frac{d^2\theta}{dy^2}$$
$$+ (e_1 - e_3)\cos(2\theta)\left(\frac{d\theta}{dy}\right)^2 + ze(n_+ - n_-) = 0 \quad (8)$$

Further, electrochemical potential for a system with dissolved ions, determined using $\mu_\pm = \frac{\delta f_{tot}}{\delta n_\pm}$ [43], is constant in the equilibrium state leading to the ionic distribution given by [48,49]

$$n_\pm = \frac{n_0 \exp(\mp ez\psi/k_B T)}{1 + \nu(\cosh(ez\psi/k_B T) - 1)} \quad (9)$$

where $\nu = 2n_0 a^3$ denotes the steric factor; $n_0$ being the number density of ions in the bulk reservoir. Substituting the value of $n_\pm$ into equation (8), we obtain the modified Poisson-Boltzmann equation for the nematic phase. The boundary condition for the potential distribution originates from the fact that the surface charge density induced at the upper substrate surface $y = h$ is equal and opposite to the cumulative net charge within the half of the fluid domain, and is given as $\sigma_w(y=1) = -\int_0^h \rho_e dy = \int_0^h 2zen_\infty \sinh\left(\frac{ze\psi}{k_B T}\right) dy$ where $\rho_e = e\sum_i z_i n_i = ez(n_+ - n_-)$ represents the net charge density. A similar argument may be used to derive the condition applicable for the lower boundary wall. Equation (8) depicts a highly intricate and coupled relationship between the potential variation and director orientation. Using the classical Poisson-Boltzmann model would incorrectly predict the potential variation due to trivially disregarding the effects of nematic director on the ionic distribution. The director orientation, in turn, is affected by the nematic elastic, dielectric and flexoelectric energies from the two individual field components and the fluid flow within the confinement, which must also be ascertained to obtain the final potential distribution. Towards this we proceed to formulate the equations governing the director orientation and electroosmotic flow.

**Electro-Nematodynamic Governing Equations**

The rod-shaped nematic director alignment distribution due to the cumulative effects of elastic, dielectric, flexoelectric energies and the fluid flow is governed by the form proposed in the Leslie-Ericksen theory which is obtained from the rate of work hypothesis for the moments involved [11]. Based on the definition of the director **n** and velocity **V** as given above, the governing equation for the angular momentum of the director reads,

$$\left(K_1 \sin^2(\theta) + K_3 \cos^2(\theta)\right)\frac{d^2\theta}{dy^2} + (K_1 - K_3)\sin(\theta)\cos(\theta)\left(\frac{d\theta}{dy}\right)^2 + \left(\alpha_3 \sin^2(\theta) - \alpha_2 \cos^2(\theta)\right)\frac{du}{dy}$$
$$+\varepsilon_0 \varepsilon_a \left[\frac{1}{2}\sin(2\theta)\left(E_1^2 - \left(\frac{d\psi}{dy}\right)^2\right) - E_1 \frac{d\psi}{dy}\cos(2\theta)\right] - \frac{1}{2}(e_1 - e_3)\sin(2\theta)\frac{d^2\psi}{dy^2} = 0 \quad (10)$$

where $\alpha_i$ ($i$=1 to 6) are the Leslie viscosities. Here we consider strong anchoring of the director near the bounding surfaces which gives the boundary equation in the form $\theta(y = -h) = \pi/2$ and $\theta(y = h) = m\pi + \pi/2$, where $m$, an integer, denotes the directional winding number [50]. Here $m\pi$ signifies the angle of rotation of the directors as traverses the channel width from $z = -h$ to $z = +h$. It must be noted that since the longitudinal axis of the rod-shaped molecules have no preferred orientation, one cannot distinguish between **n** and $-$**n** [11,13]. Thus, a rotation of the director by a multiple of $\pi$ near the boundary is absolutely allowed [50] for a planar configuration of the present type. The essence of using such a bounding condition for $\theta$ will be elaborated while discussing the existence of multiple solutions, which are integral part of the solution process considerations when concerning liquid crystal flows.

An application of an external electric field induces a flow field which needs to be determined in order to obtain the director configuration of the nematic crystals under dynamic equilibrium condition. Towards this, we appeal to the Leslie-Ericksen that governs the general flow of a nematic fluid with a proposed constitutive behavior having the deviatoric stress relation as equations [11,13]

$$\sigma_{ij} = \alpha_1 n_i n_j A_{kp} n_k n_p + \alpha_2 n_j N_i + \alpha_3 n_i N_j + \alpha_4 A_{ij} + \alpha_5 n_j A_{ik} n_k + \alpha_6 n_i A_{jk} n_k \quad (11)$$

where $N_i = Dn_i/Dt - W_{ij}n_j$ is the co-rotational vector representing the rate of change of director with respect to the background fluid while $A_{ij}$ and $W_{ij}$ is the symmetric and anti-symmetric part of the strain tensor $\nabla \mathbf{V}$. In our case for steady, electroosmotically driven flow of a nematic through a narrow confined cell with no-slip boundary conditions $u(y = \pm h) = 0$ and an electroosmotic body force, the linear momentum balance equation in the flow direction reduces to

$$\frac{d}{dy}\left(\eta(\theta)\frac{du}{dy}\right) = -\rho_e E_1 \tag{12}$$

where $\rho_e E_1$ is the electroosmotic body force density and the position-dependent apparent nematic crystals viscosity is given by $\eta(\theta) = \eta_1 \sin^2\theta + \eta_2 \cos^2\theta + \eta_{12} \sin^2\theta \cos^2\theta$. Here the viscosity parameters $\eta_1, \eta_2$ and $\eta_{12}$ are the Miesowicz viscosities related to the Leslie viscosities by the relations $\eta_1 = \frac{\alpha_3 + \alpha_4 + \alpha_6}{2}$, $\eta_2 = \frac{-\alpha_2 + \alpha_4 + \alpha_5}{2}$ and $\eta_{12} = \alpha_1$ [11]. The set of coupled equations (8), (10) and (12) governs the electroosmosis of nematic LC, which needs to be solved with appropriate bounding conditions to obtain our results. Before proceeding to the result section, we derive an equivalent dimensionless forms of the governing equations and the boundary conditions thereby obtaining a more general form that regulate such flows.

**Non-dimensionalization of the Governing Equations**

To obtain the corresponding dimensionless forms for governing equations, we choose the scaling references as follows: $\bar{y} = y/h$, $\bar{\psi} = ze\psi/k_B T$, $\bar{u} = u/u_{ref}$, $\bar{E}_x = E_x/E_{x,ref}$, $\bar{E}_y = E_y/E_{y,ref}$; the reference values will be rationalized subsequently. With the present dimensionless considerations and employing equation (9), we obtain the modified Poisson-Boltzmann (8) for nematic fluids as

$$\left(1 + \frac{\varepsilon_a}{\varepsilon_\perp}\cos^2(\theta)\right)\frac{d^2\bar{\psi}}{d\bar{y}^2} - \left(\frac{\bar{E}_x \cos(2\theta)}{p_1} + \sin(2\theta)\frac{d\bar{\psi}}{d\bar{y}}\right)\frac{d\theta}{d\bar{y}}$$
$$+ A_4\left\{\sin(2\theta)\frac{d^2\theta}{d\bar{y}^2} + 2\cos(2\theta)\left(\frac{d\theta}{d\bar{y}}\right)^2\right\} - \frac{\sinh(\bar{\psi})}{\bar{\lambda}^2\left(1 + \nu(\cosh(\bar{\psi}) - 1)\right)} = 0 \tag{13}$$

The corresponding dimensionless boundary condition at both the charged surface as obtained employing the charge neutrality condition is $\bar{\sigma}_w = \int_0^1 \frac{1}{\bar{\lambda}^2} \sinh(\bar{\psi}) d\bar{y}$ where $\bar{\sigma}_w = \frac{zeh\sigma}{\varepsilon_0 \varepsilon_\perp k_B T}$ denotes the dimensionless surface charge density, $\bar{\lambda} = \frac{\lambda}{h} = \sqrt{\frac{\varepsilon_0 \varepsilon_\perp k_B T}{2z^2 e^2 n_0 h^2}}$; $\lambda$ being the dimensional Debye screening length and $p_1 = \frac{k_B T}{E_{c1} h z e}$ and $A_4 = \frac{ze(e_1 - e_3)}{2(\varepsilon_0 \varepsilon_\perp) k_B T}$. The dimensionless form for the equation governing the director orientational distribution is then obtained in the form

$$\left(\sin^2(\theta) + \kappa \cos^2(\theta)\right) \frac{d^2\theta}{d\bar{y}^2} + (1-\kappa) \sin(\theta)\cos(\theta)\left(\frac{d\theta}{d\bar{y}}\right)^2 + m\left(\sin^2(\theta) - \frac{\alpha_2}{\alpha_3}\cos^2(\theta)\right)\frac{d\bar{u}}{d\bar{y}}$$
$$+q\left[(\bar{E}_x^2 - p^2 \cdot \bar{E}_y^2)\sin(2\theta) + 2 \cdot p \cdot \bar{E}_x \cdot \bar{E}_y \cos(2\theta)\right] + w\sin(2\theta)\frac{d\bar{E}_y(\bar{y})}{d\bar{y}} = 0 \quad (14)$$

where the dimensionless parameters have the following form: $\kappa = K_3/K_1$, $\bar{\alpha}_3 = \alpha_3/\eta_{ref}$, $E_{2,ref} = \frac{E_2}{\sigma_w/\varepsilon_0 \bar{\varepsilon}}$, $\bar{\alpha}_2 = \alpha_2/\eta_{ref}$, $q = \frac{\varepsilon_0 \varepsilon_a E_{c1}^2 h^2}{2K_1}$;, $m = \frac{u_{ref} h \eta_{ref}}{K_1}$ $p = \frac{\sigma_w}{\varepsilon_0 \bar{\varepsilon} E_{c1}}$ and $w = \frac{e_1 - e_3}{2K_1}\left(\frac{\sigma h}{\varepsilon_0 \bar{\varepsilon}}\right)$ with $\bar{\varepsilon} = \frac{\varepsilon_\parallel + 2\varepsilon_\perp}{3}$. The viscosity reference has been chosen as $\eta_{ref} = \alpha_4/2$ which signifies the corresponding Newtonian counterpart of the nematic viscosity as noted from deviatoric stress equation (11) while the velocity reference $u_{ref}$ can now finally be obtained from the flow equation whose dimensionless form is given by

$$\frac{d}{d\bar{y}}\left(\bar{\eta}(\theta)\frac{d\bar{u}}{d\bar{y}}\right) = \sinh(\bar{\psi})\bar{E}_1 \quad (15)$$

where $\bar{\eta}(\theta) = \frac{\eta(\theta)}{\eta_{ref}} = \sin^2\theta + (\eta_2/\eta_{ref})\cos^2\theta + (\eta_{12}/\eta_{ref})\sin^2\theta\cos^2\theta$ is the local viscosity and the velocity scale reads $u_{ref} = \frac{2zen_0 E_{c1} h^2}{\eta_{ref}}$. Before proceeding directly to the solution of the coupled PBLE (Poisson-Boltzmann-Leslie-Ericksen) equations for the nematics, we make certain subtle non-trivial observation from our dimensionless governing equations and comment on the velocity scale chosen for the present study. In the absence of any flow velocity implying absence of an applied axial electric field, we have $du/dy = 0$ and $E_1 = 0$, thereby the reduced form of equation (10) gives the static configuration of the nematic director due to an spontaneous EDL formation. It is further noteworthy that in the limit $E_1 = 0$ and $v = 0$ our set of governing equation reduces to the steady state limit form as provided in [18,26,37,38] where ionic inclusions have also been considered. Nevertheless, the presence of an induced EDL due to surface electrochemistry, in addition to an axial field dominated director configuration and flow actuation, have not been previously looked into in the considerations of ER behavior of nematic fluids.

## Results and Discussion

In this section, we first explore the existence of multiple solution for the present study and focus on the stability of such solutions through an entropic analysis before elaborating on the corresponding electroosmotic flow characteristics for nematic LCs. We solve the coupled dimensionless equations (13), (14) and (15) with their appropriate boundary conditions. For a representative case, we have selected the nematic MBBA (*N*-(4-Methoxybenzylidene)-4-butylaniline) for our calculation whose properties are detailed in **Table 1**. The induced surface charge density is varied between $10^{-5} - 10^{-2}$ Cm$^{-2}$ while a concentration of ionic impurities is considered in order of $10^{-3}$ m-M with ionic shell size in order of $10^{-9}$ m [51–53] which results in a steric factor $v \sim 10^{-3}$ [52,54,55] and the order of dimensionless charge density in the range of $\sim 10^2 - 10^4$. Here $E_{1,ref}$ is considered in the order of $10^6$ *kV/mm*; a scale close to the Freedericksz transition electric field $E_{c1}$ that is defined as the threshold electric field above which deformations in the nematic director is observed [11,14]. An interesting aspect of MBBA is its negative dielectric anisotropy due to which the molecules tend to orient themselves perpendicular to the electric field. Such an aspect in presence of ionic impurities has not been hitherto rigorously focused in the past literature which further motivates us to consider MBBA for our present study.

| Property | Property Value | Unit | Property | Property Value | Unit |
|---|---|---|---|---|---|
| Splay elastic constant ($K_1$) | 5 | pN | $\alpha_5$ | 0.0779 | Pa-s |
| Bend elastic constant ($K_3$) | 7.5 | pN | $\alpha_6$ | -0.0336 | Pa-s |
| Viscosity coefficient ($\alpha_1$) | -0.0181 | Pa-s | Flexoelectric coefficient ($e_1 - e_3$) | 14 | pC/m |
| $\alpha_2$ | -0.1104 | Pa-s | Dielectric permittivity (relative) $\varepsilon_\parallel$ | 4.7 | — |
| $\alpha_3$ | -0.001104 | Pa-s | $\varepsilon_\perp$ | 5.4 | — |
| $\alpha_4$ | 0.0852 | Pa-s | | | |

**Table 2:** Symbols, magnitudes and units of the MBBA nematic properties used for the present study.

Multiplicity of solution is an integral part of nematic LC solution due to different orientational possibilities of the director arrangement [7,50,56–58]. These multiple solutions can be physically realized by varying the initial condition of the director arrangement or flow actuation mechanism or may be a intermediate state in the path of the eventual orientational arrangement in a flow. However, it must be emphasized that most of these multiple solutions are unstable and a careful analysis must be performed in order to distinguish the most stable one. In this section, we depict the multiple form of the solutions obtained and subsequently carry out an entropic analysis for the corresponding solutions to understand their degree of stability.

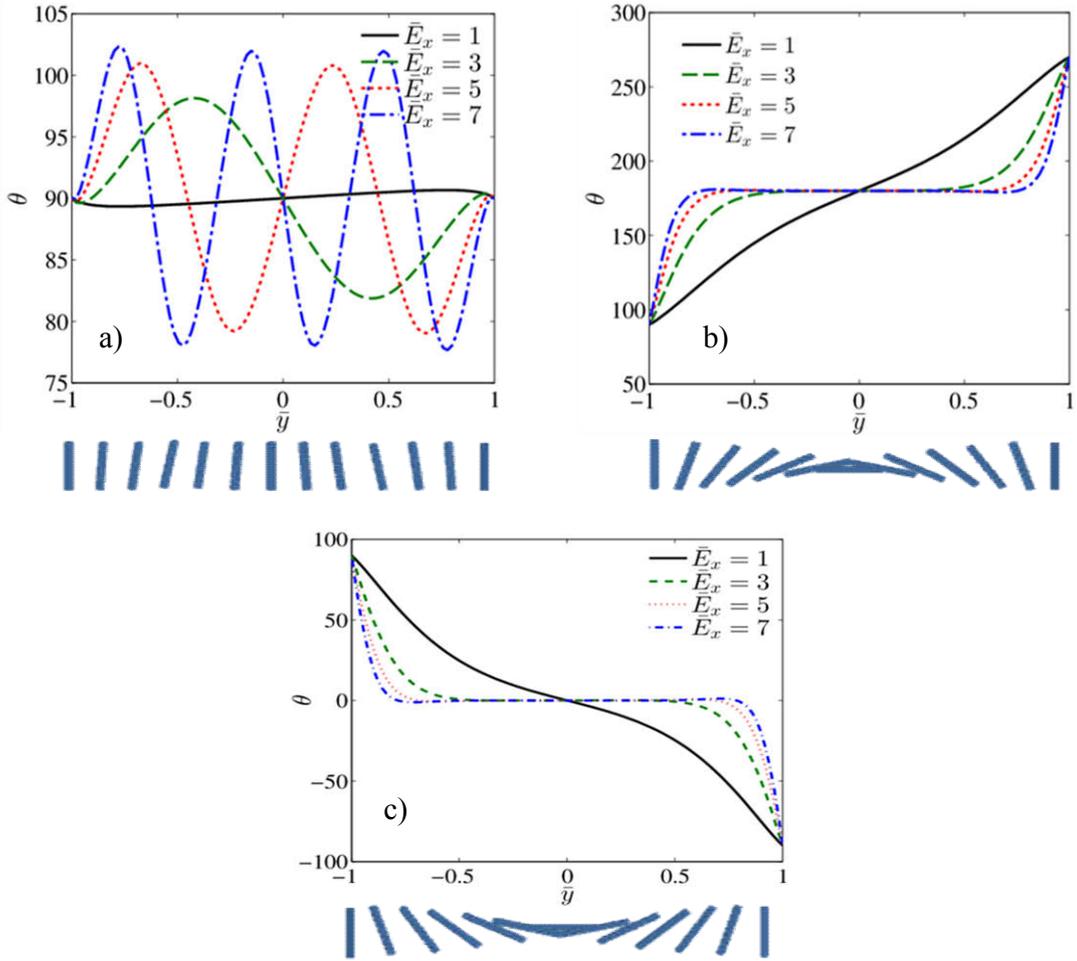

**Fig 2.** Depicts the multiple arrangements of the director using the director orientation angle $\theta$ for different values of the axial electric field $\bar{E}_x$ for $\bar{\lambda}=0.1$ and $\bar{\sigma}_w=-2000$.

Figure 2 depicts the various director arrangements $\theta$ obtained for the electroosmotic flow for different values of $\bar{E}_x$ with $\bar{\lambda}=0.1$ and $\bar{\sigma}_w=-2000$. The three sub-plots in figure 2 are drawn for different boundary conditions with $m=0$, $m=1$ and $m=-1$, respectively. The various choice of $m$ determines the direction of rotation allowed on the director for the electroosmotic flow. However, these different boundary conditions are in fact similar for director orientation at the boundaries since a rotation of $\pi$ on the director does not change the boundary condition definition owing to the fact that **n** and $-$**n** are not distinguishable. Nevertheless, such a differently invoked definition at the boundaries result in a completely dissimilar director distribution within the channel, and thus, depict possible solutions to the problem among which one may be the most stable one. Figure 2a shows that as the axial electric field strength is increased, the periodic fluctuation characteristics of the director is brought into existence wherein the director tumbles about $\theta = \pi/2$. With higher field, the periodic amplitude as well as the frequency is increased which is attributable to the a coupled, competing and intricate influence of the potential distribution and resulting flow rate variation which tries to keep the nematic in the planar arrangement while the external field which tries to rotate the molecular rods perpendicular to it due to the negative dielectric anisotropy. Figure 2b shows another possible solution wherein we observe a clockwise rotation of the director as it moves from the lower plate to the upper one while figure 2c displays an exact opposite result. One important consequence of negative dielectric anisotropy is that for high fields the director are oriented perpendicular to the field direction, which is clearly observed in these figures. This characteristic of the director alignment qualitatively hint towards a higher stability for solution in figure 2b and figure 2c as compared to figure 2a. Nonetheless, we address a fundamental quantitative approach to recognize the stable solution mode.

*Entropy generation rate*

Among the solutions obtained above in the present analysis, we must reiterate the fact that most of them are unstable and a fundamental investigation is demanded in order to distinguish them. Towards this, we appeal to the entropy generation rate for a nematic liquid undergoing an electroosmotic flow and proceed to evaluate the same for the different solutions obtained from the governing equations. The total entropic generation rate for an electroosmotic flow of a nematic LC may be evaluated from the expression [13,59,60]

$$T \dot{S}'' = \int \left( \bar{\sigma}_{ij} \frac{\partial v_i}{\partial x_j} + \mathbf{h} \cdot \frac{D\mathbf{n}}{Dt} + \frac{E_1^2}{\sigma_0} \right) d^3 \mathbf{r} \tag{16}$$

where $\bar{\sigma}_{ij}$ is the viscous stress tensor (eq. (11)) with $V_{ij} = \frac{1}{2}\left( \frac{\partial v_i}{\partial x_j} + \frac{\partial v_j}{\partial x_i} \right)$ representing the symmetric part of the velocity gradient tensor and $N_i = \dot{n}_i - \frac{1}{2}\left( \frac{\partial v_i}{\partial x_j} - \frac{\partial v_j}{\partial x_i} \right) n_j$ being the rate of change of director with respect to the background fluid. Here $\mathbf{h} = \frac{\delta F}{\delta n_m}$ denotes the molecular field for the nematic liquid and $\sigma_0$ is the electrical resistivity of the liquid. In the steady state with given director alignment, equation (16) reduces to

$$T\dot{S}'' = \int_{-h}^{h}\left(\eta(\theta)\left(\frac{du}{dy}\right)^2 + \frac{E_1^2}{\sigma_0}\right)dy \qquad (17)$$

We report the rescaled dimensionless entropic generation rate per unit area as defined by

$$\bar{\dot{S}}'' = \frac{\dot{S}_r''}{\dot{S}_{ref}''} \qquad (18)$$

in **Table 2** where $\dot{S}_r'' = \dot{S}'' - \frac{E_1^2}{\sigma_0}$ and $\dot{S}_{ref}'' = \frac{u_{ref}^2 \eta_{ref}}{hT}$.

| Solution No. | Boundary Conditions | Entropy Generation rate $\left(\dot{S}_r''\right)$ | Rotation | Center alignment $(\theta_m)$ |
|---|---|---|---|---|
| 1 | $\theta\big|_{\bar{y}=-1} = \pi/2$, $\theta\big|_{\bar{y}=1} = \pi/2$ | 20.7768 | — | $\pi/2$ (Parallel to flow) |
| 2 | $\theta\big|_{\bar{y}=-1} = \pi/2$, $\theta\big|_{\bar{y}=1} = 3\pi/2$ | 15.0598 | CLOCKLWISE | $\pi$ (Perpendicular to flow) |
| 3 | $\theta\big|_{\bar{y}=-1} = \pi/2$, $\theta\big|_{\bar{y}=1} = -\pi/2$ | 8.2262 | COUNTER-CLOCKLWISE | 0 (Perpendicular to flow) |

**Table 2:** Details of the existence of multiple solutions, their corresponding dimensionless entropies ($\bar{s}$) and solution characteristics for different boundary conditions for specific choice of parameters as: $\bar{E}_1 = 7$; $\bar{\lambda} = 0.1$ and $\bar{\sigma}_w = -2000$.

To explore the most stable configuration we resort to entropy production principle [57] which state that the system will adopt the configuration which minimizes the entropy production. Following table 2, we realize that *solution 3* projects the most stable case and is most likely to be obtained in experimental situations. In fact, we have seen that such stability persists across the range of low to high applied electric field. We, therefore, proceed with the conditions of *solution 3* in the rest of our study depicting the potential distribution, director orientation and electroosmotic velocity of nematic MBBA liquid crystal. We further focus on the observed electrorheological phenomena, volume throughput and existence of normal stress difference attributed to the liquid's elastic nature.

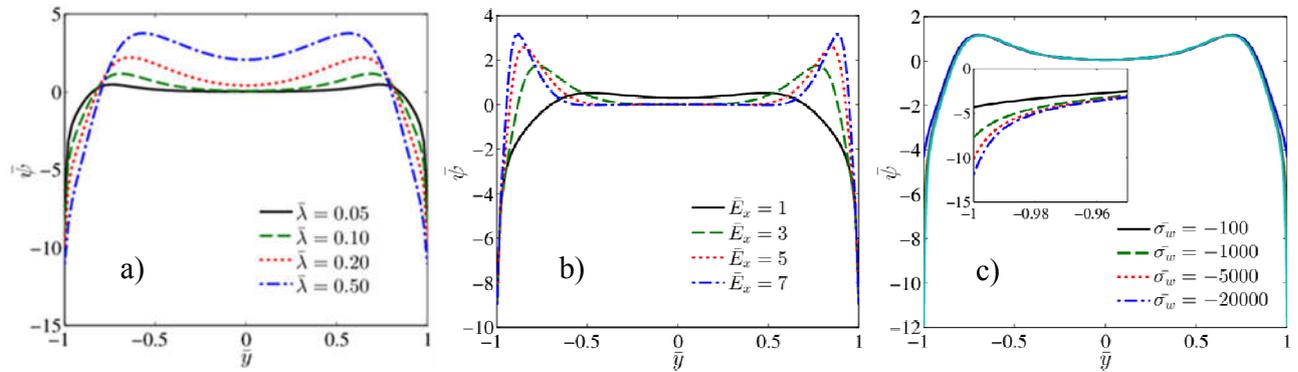

**Fig 3.** Depicts the variation of the potential distribution profile as a function of the channel transverse direction for a) different values of the inverse dimensionless Debye screening length $\bar{\lambda}$ with $\bar{E}_1 = 7$ and $\bar{\sigma}_w = -2000$; b) different values of dimensionless axial applied field $\bar{E}_1$ with $\bar{\lambda} = 0.1$ and $\bar{\sigma}_w = -2000$; and c) different values of the wall charge density $\bar{\sigma}_w$ with $\bar{E}_1 = 7$ and $\bar{\lambda} = 0.1$.

Figure 3 shows the dimensionless potential distribution profile $\bar{\psi}$ as a function of the transverse direction $\bar{y}$ for different values of $\bar{\lambda}$, $\bar{E}_1$ and $\bar{\sigma}_w$. A general aspect that is observed from the figures is the prediction of charge inversion effect especially for higher values of $\bar{\lambda}$ and $\bar{E}_1$. Charge inversion is a common phenomenon in electrokinetic studies [61,62] wherein the charged surface binds the counterions so strongly that there occurs regions of higher concentration of coions which can advect more freely compared to the counterions that remains closely bound to the surface or in some cases the surface charge itself effectively inverses sign.

In figure 3a, we observe that for a larger $\bar{\lambda}$ implying thicker EDL, the phenomena of charge inversion enhances. This may be attributed to the effect of transverse field which can penetrate deeper into the channel mid-plane. Figure 3b depicts the potential distribution profile for different values of dimensionless axial electric field strength $\bar{E}_1$. In sharp contrast to Newtonian medium, a non-intuitive behaviour is showcased in this case where an axial field modifies the potential distribution, and thereby, the charge number density profiles within the channel. It can also be seen that higher field induces larger charge inversion phenomenon which is attributed to the non-linear interplay between higher director deformation and larger electrical body force. Figure 3c shows the variation of the potential distribution for different surface charge density. It clearly shows that higher the induced surface charge, higher is the resulting surface potential (as seen in the figure inset), and consequently, a larger transverse electric field is induced. However, the potential distribution does not drastically vary in the diffuse layer and the bulk since the counterion-coion layers screen the surface charge while the axial electric field dominate in these regions. It must be noted here that the maximum dimensionless induced potential is in the order of 10 which transpires to 0.25 volts in dimensional values which is frequently encountered in electrokinetic studies [30,54]. Figure 3 in general shows the intrinsic effect of various parameters on potential distribution which further depends non-linearly on the director alignment across the channel that we shall determine in the next section.

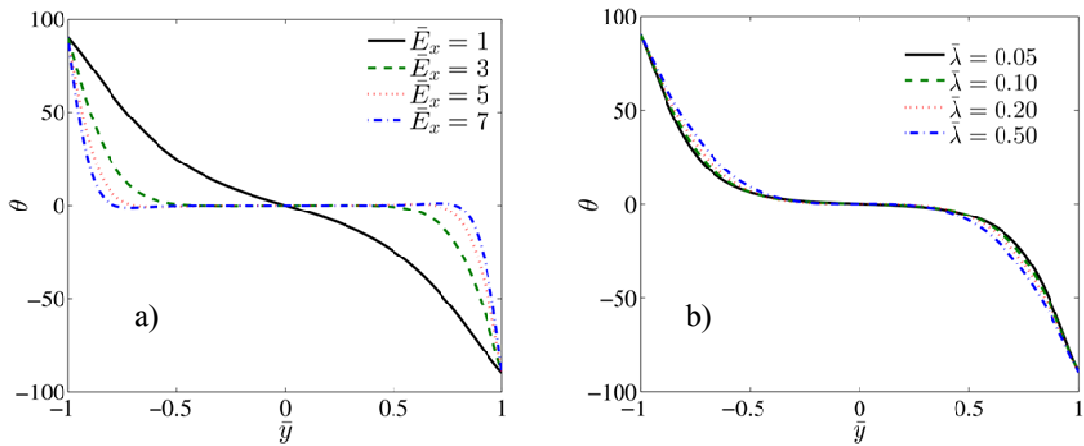

**Fig 4.** Depicts the variation of the director configuration profile as a function of the channel transverse direction for a) different values of dimensionless axial applied field $\bar{E}_1$ with $\bar{\lambda} = 0.1$

and $\bar{\sigma}_w = -2000$ ;; and b) different values of the inverse dimensionless Debye screening length $\bar{\lambda}$ with $\bar{E}_1 = 7$ and $\bar{\sigma}_w = -2000$.

Figure 4 shows the variation of director alignment across the channel width for various values of the applied electric field and dimensionless Debye screening length. For both the solution sub-plots, it can clearly be seen that the orientation of the director at the channel centerline region is perpendicular to the axial electric field direction which is attributable to the negative dielectric anisotropy value of MBBA used in our study. In figure 4a, it is seen that a higher electric field augments the region span within the bulk where the director is oriented near orthogonally to the applied field direction which is a generic director alignment feature for negative $\varepsilon_a$ as explained above. Since the induced alignment of the director due to the electric field is normal to the flow direction, a higher viscosity is expected in this case which gives rise to the *electrorheological effect* (ER) in the present study. ER effect have been previously demonstrated in nematic flow studies predominantly with transverse imposed electric field [5,7,20,35]. Here we showcase a scenario wherein an axial electric field which effectively drives the flow also contributes to an induced ER effect (as will be elaborated in figure 6). figure 4b depicts the director alignment profile for variation of different $\bar{\lambda}$ where an interesting aspect of higher penetration of the transverse field into the channel centerline can be observed. It is seen that for larger $\bar{\lambda}$, depicting higher Debye length, the span of the director alignment orthogonal to the axial field reduces and the planar arrangement dominates in greater part of the channel. However, such a dependence on the dimensionless Debye length is small, and therefore, the corresponding viscosity variation due to $\bar{\lambda}$ variation, as will be demonstrated later, is rather less. The effect of the surface charge density $\bar{\sigma}_w$ variation on the director alignment has been found to be negligible and therefore, for the sake of brevity, has not been included in the depiction.

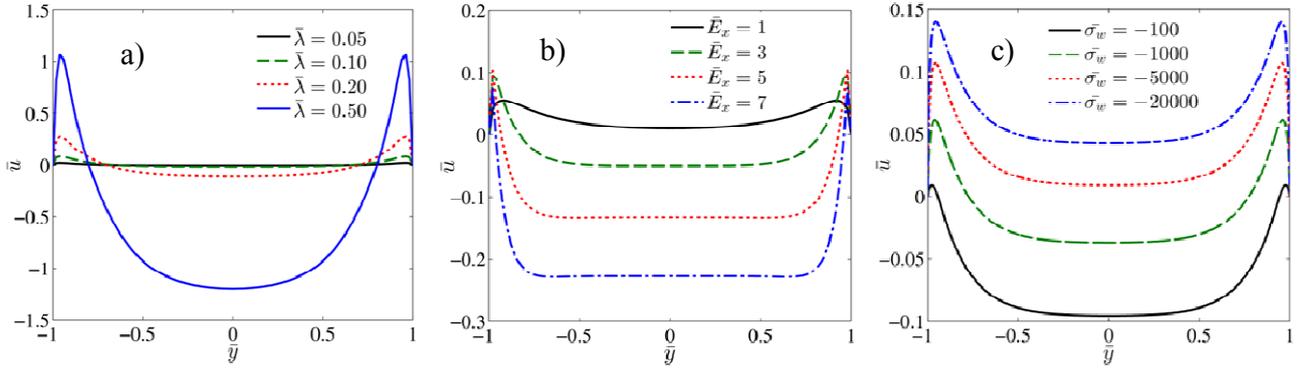

**Fig 5.** Depicts the variation of the velocity field profile for different values of a) the inverse dimensionless Debye screening length $\bar{\lambda}$ with $\bar{E}_1 = 7$ and $\bar{\sigma}_w = -2000$; b) the dimensionless axial applied field $\bar{E}_1$ with $\bar{\lambda} = 0.1$ and $\bar{\sigma}_w = -2000$; and c) different values of the wall charge density $\bar{\sigma}_w$ with $\bar{E}_1 = 7$ and $\bar{\lambda} = 0.1$.

Figure 5 depicts the velocity profile in the confined conduit for different values of a) $\bar{\lambda}$; b) $\bar{E}_1$ and c) $\bar{\sigma}_w$. Before proceeding to elaborate the velocity profile characteristics, a subtle feature must be discussed. We observe a general pattern of reverse flow or backflow phenomena which is again a common occurrence in electroosmotic flows [63,64], for high axial field and dimensionless screening length. Backflow is the phenomenon wherein the resultant flow direction is opposite to what is expected due to counterion-axial field interaction and is attributable to the charge inversion as discussed above. Second, is the suitability of the choice of the reference velocity scale, which differs from the classical Smoluchoski scale for Newtonian fluids, may be justified from the observed dimensionless velocity magnitudes that remains in order of unity. Figure 5a depicts the variation of the velocity profile across the channel width for different $\bar{\lambda}$ values. We find that for higher $\bar{\lambda}$, that corresponds to a stronger charge inversion, there is an enhanced flow reversal. As the Debye screening length reduces, the flow reversal also tends to vanish. A similar scenario is also observed in figure 5b, where the flow reversal, corresponding to the charge inversion, gets enhanced with higher axial electric field. For low fields, however, the flow reversal is absent and the flow profile almost corresponds to a Newtonian case. This observation may be reasoned on the fact that at low axial field, the director remains aligned parallel to the channel wall due to planar boundary conditions and the induced transverse field. In fact, an unambiguous comprehension may be attempted from figure 5c where we clearly observe that for higher surface charge density, there is no flow reversal since at such high surface charge density, implying a stronger transverse field, the director orientation remain parallel to the wall plane and a rheology and flow conditions tending towards Newtonian case is more dominant. However, at low surface charge, the director orientation is dominated by the axial field and due to the coupled interplay between director alignment and potential distribution a charge inversion results associated with a strong flow reversal.

*Electrorheological Viscoelasticity*

An appealing aspect that comes forward in nematic crystal flows is the local viscosity variation $\eta(\theta)$ and existence of a first normal stress difference $N_1 = \sigma_{xx} - \sigma_{yy}$ due to its elastic property. The form for the local viscosity variation, as stated above, reads $\bar{\eta}(\theta) = \frac{\eta(\theta)}{\eta_{ref}} = \sin^2\theta + (\eta_2/\eta_{ref})\cos^2\theta + (\eta_{12}/\eta_{ref})\sin^2\theta\cos^2\theta$ which fluctuates due to the coupled influence of the applied and induced electric field as well as the nematic director distribution. An average viscosity may be defined as $\langle\bar{\eta}\rangle = \frac{1}{2}\int_{-1}^{1}\bar{\eta}(\theta(\bar{y}))d\bar{y}$ that denotes the apparent viscous effect observed in the electroosmotic flow. Since the presence of electric field enhances the average viscosity of the nematic fluid when compared to only planar arrangement case, an electrorheological (ER) effect usually gets associated with these flows. One strong advantage of using nematic LCs as ER fluids over ER suspensions is the absence of suspended particles in the nematic medium apart from the fact that the individual molecules are of very small size($\sim 20 \overset{o}{A}$). These properties prevent agglomeration, sedimentation and abrasion problems making them appreciable for microfluidic or even nanofluidic flow applications [8]. Besides such rheological variations, a first normal stress difference exists which can be measured suing the form $\bar{N}_1(\theta(y)) = -\sin(2\theta)(\bar{\alpha}_1\cos(2\theta) + \bar{\alpha}_2 + \bar{\alpha}_3)\frac{d\bar{u}}{d\bar{y}}$ where $\bar{\alpha}_i = \frac{\alpha_i}{\eta_{ref}}$ are the dimensionless Leslie viscosities and $\bar{N}_1 = \frac{N_1}{N_{1,ref}}$ is the dimensionless normal stress difference with $N_{1,ref} = \frac{\eta_{ref} u_{ref}}{2h}$.

Following the similar averaging procedure, an averaged first normal stress difference may be defined as $\langle \bar{N}_1 \rangle = \frac{1}{2}\int_{-1}^{1} \bar{N}_1(\theta(\bar{y}))d\bar{y}$, a quantity measurable in experimental studies.

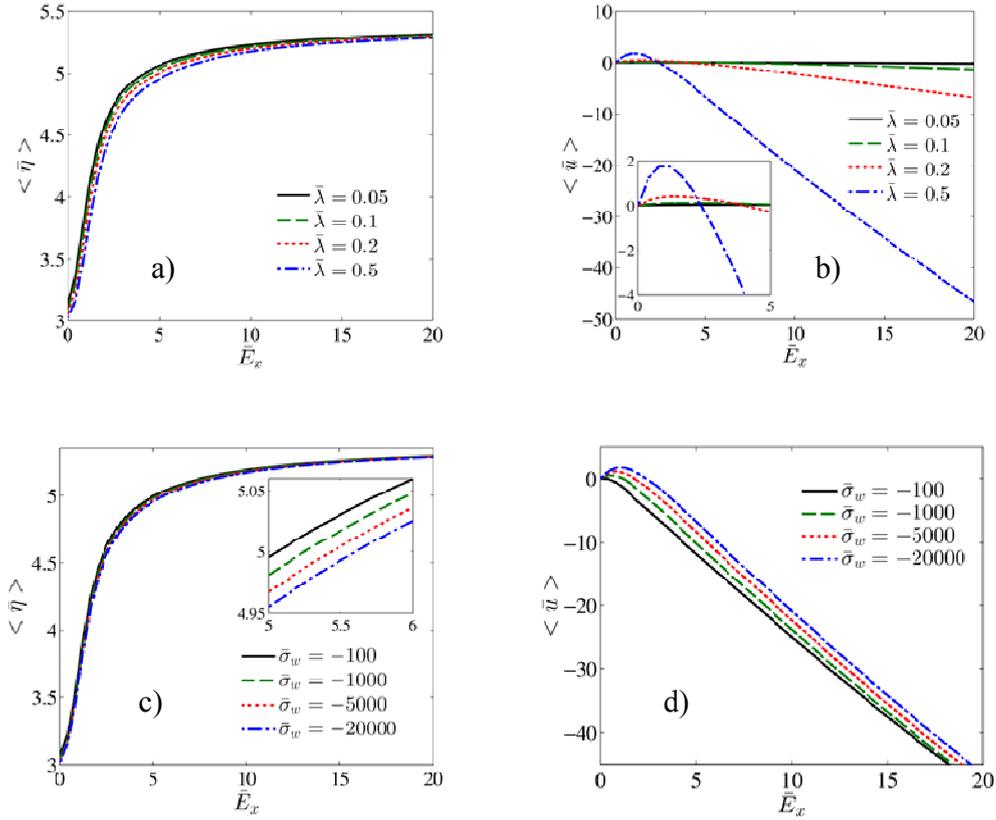

**Fig 6.** Depicts the variation of a) the average viscosity and b) the average velocity as a function of the axial field $\bar{E}_1$ for different values of the inverse dimensionless Debye screening length $\bar{\lambda}$ with $\bar{\sigma}_w = -20000$; c) the average viscosity and d) the average velocity as a function of the axial field $\bar{E}_1$ for different values of dimensionless charge density $\bar{\sigma}_w$ with $\bar{\lambda} = 0.5$.

Figure 6a and figure 6b depicts the variation of the average viscosity $\langle \bar{\eta} \rangle$ and the corresponding volume flow rate, defined as $\langle \bar{u} \rangle = \frac{1}{2}\int_{-1}^{1} u(\bar{y})d\bar{y}$, as a function of the axial electric field for different values of dimensionless Debye screening length $\bar{\lambda}$ with $\bar{\sigma}_w = -2000$. As discussed earlier, it can be verified from figure 6a that as the axial field is enhanced, the electrorheological effects gets prominent owing to higher apparent viscosity. It is also seen that as the $\bar{\lambda}$ decreases, the influence of the transverse field to orient the director in line with the velocity weakens thereby enhancing the average viscosity. Figure 6b shows that the for lower applied field a positive volume flow rate is observed in the channel consistent with the discussion in figure 5b. The volume flow rate, however, reverses sign at higher applied field due to flow reversal. The effect of the flow reversal is furthermore augmented in cases of higher values of $\bar{\lambda}$ consistent with figure 5a. Figure 6c and figure 6d depicts the variation of the average viscosity $\langle\bar{\eta}\rangle$ and the corresponding volume flow rate $\langle\bar{u}\rangle$ as a function of the axial electric field for different values of dimensionless Debye screening length $\bar{\sigma}_w$ with $\bar{\lambda}=0.1$. The trend of $\langle\bar{\eta}\rangle$ variation with $\bar{E}_1$ remains similar; however, it is seen the surface charge density has negligible influence on the ER effect. The inset displays an enlarged view of the plot wherein we find that higher $\bar{\sigma}_w$ shows lower ER effect attributable to the induced stronger transverse field. Figure 6d shows that the average velocity is positive for lower axial field but reverses sign as the field is increased. It is also apparent that as $\bar{\sigma}_w$ is increased, the average flow velocity due to a lower average viscous hindrance as demonstrated before.

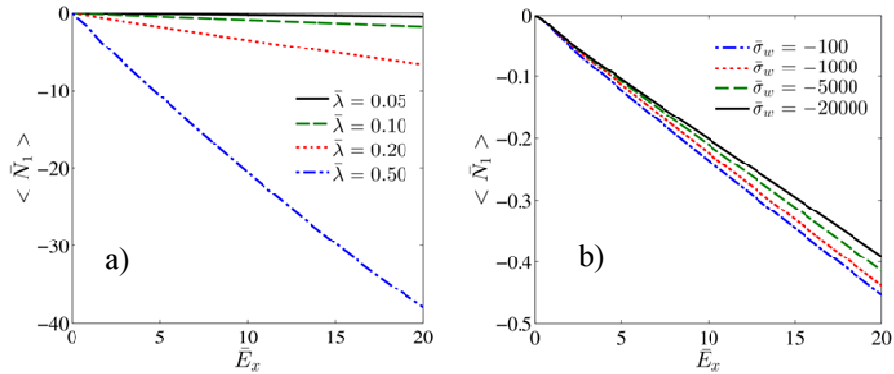

**Fig 7.** Depicts the variation of the average normal stress difference $\langle\bar{N}_1\rangle$ as a function of the axial field $\bar{E}_1$ for different values of the a) inverse dimensionless Debye screening length $\bar{\lambda}$ with $\bar{\sigma}_w=-100$ and b) for different values of dimensionless charge density $\bar{\sigma}_w$ with $\bar{\lambda}=0.05$.

Figure 7 depicts the average normal stress difference $\langle \bar{N}_1 \rangle$ as a function of the electric field for different values of dimensionless Debye screening length and surface charge density. The average normal stress, which is absent in Newtonian fluids, exists within the nematic LC due to its intrinsic elastic characteristics and molecular orientation. In general, a negative normal stress difference is observed, implying that the normal force tends to join the two confining walls, which is in sharp contrast to shear flow of nematic LCs in presence of a transverse field where both positive and negative normal stress difference may be observed. It is also apparent that the magnitude of the normal stress difference $\langle \bar{N}_1 \rangle$ linearly increases with the applied axial electric. Figure 7a and figure 7b shows that as $\bar{\lambda}$ and $\bar{\sigma}_w$ is increased, respectively, a stronger first normal stress difference $\langle \bar{N}_1 \rangle$ is encountered. This is attributed to the enhanced effect of the transverse field on the flow which augments the $\langle \bar{N}_1 \rangle$. However, the variation of $\langle \bar{N}_1 \rangle$ with $\bar{\sigma}_w$ is comparatively lower than due to $\bar{\lambda}$; although, in general, an attractive trend between the confining walls is expected in generality from the present study.

## Conclusion

Electro-nematodynamic study employing an electrical field driven flow in presence of ionic species in the liquid sample has been studied in the present work. Due to the presence of an EDL, a transverse field gets induced which along with the axial actuating field intrinsically influences the resultant flow rheology. A fundamental free energy for the nematic LC with dissolved ions is developed and a variational approach is employed to derive the governing equation for potential distribution, director alignment and flow velocity. We further explore the existence of multiple solutions for the LC flows and quantitatively commented on the stability of such flows with entropic generation rate evaluation for a particular nematic MBBA. Proceeding with the most stable solution, the general potential distribution, director orientation and flow characteristics is depicted. Finally, we focused on the electrorheological nature and viscoelastic properties of the nematic LC. We observed that in general the ER effect gets augmented while the first normal stress difference displays a linear variation with increase in the axial applied field. We believe the present investigation will be a precursor to future experimental studies of electroosmosis of nematic LCs under electrical double layer phenomena.